\PassOptionsToPackage{usenames}{xcolor}
\PassOptionsToPackage{dvipsnames}{xcolor}

\documentclass[manuscript, sigconf, nonacm]{acmart} 

\usepackage{csquotes}
\usepackage{subcaption}
\usepackage{enumitem}
\usepackage{amsmath}
\usepackage{url}
\usepackage{balance}
\usepackage{siunitx}
\usepackage[usenames]{xcolor}

\usepackage{array}
\newcolumntype{P}[1]{>{\centering\arraybackslash}m{#1}}
\newcolumntype{L}[1]{>{\arraybackslash}m{#1}}
\usepackage{xspace}
\newcommand\ie{i.\,e.\xspace}
\newcommand\eg{e.\,g.\xspace}

\DeclareMathOperator{\logit}{logit}

\definecolor{DeepBlue}{HTML}{1A476F}
\definecolor{DeepRed}{HTML}{90353B}
\definecolor{DarkGreen}{HTML}{55752F}
\definecolor{Orange}{HTML}{E37E00}
\definecolor{BrightYellow}{HTML}{F0D048}
\definecolor{RichRed}{HTML}{C10534}
\definecolor{LightBlue}{HTML}{6098C8}
\definecolor{SoftCerulean}{HTML}{8DD3C7}
\definecolor{Coral}{HTML}{F08070}
\definecolor{SoftBlue}{HTML}{7B92A8}
\definecolor{Teal}{HTML}{2D6D66}
\definecolor{DarkYellow}{HTML}{9C8847}
\definecolor{LightBrown}{HTML}{BFA19C}
\definecolor{LightPurple}{HTML}{938DD2}

\AtBeginDocument{%
	\providecommand\BibTeX{{%
			\normalfont B\kern-0.5em{\scshape i\kern-0.25em b}\kern-0.8em\TeX}}}

\setcopyright{none}
\copyrightyear{2023}
\acmYear{2023}
\acmDOI{}

\acmConference[]{}{}{}
\acmBooktitle{}
\acmPrice{}
\acmISBN{}

\begin{document}
	
\fancyhead{}

\title{Content Moderation on Social Media in the EU:\\ Insights From the DSA Transparency Database}

\author{Chiara Drolsbach}
\email{chiara.drolsbach@wi.jlug.de}
\affiliation{
	\institution{JLU Giessen}
	\streetaddress{Licher Str.\ 74}
	\country{Germany}
}

\author{Nicolas Pröllochs}
\email{nicolas.proellochs@wi.jlug.de}
\affiliation{
	\institution{JLU Giessen}
	\streetaddress{Licher Str.\ 74}
	\country{Germany}
}
\renewcommand{\shortauthors}{Chiara Drolsbach and Nicolas Pröllochs}

\begin{abstract}
	The Digital Services Act (DSA) requires large social media providers to provide clear and specific information whenever they remove or restrict access to certain content on their platforms in the EU. These ``Statements of Reasons'' (SoRs) are collected in the DSA Transparency Database to ensure transparency and scrutiny of content moderation decisions of the providers of online platforms. 
	In this work, we empirically analyze this unique data source and provide an early look at content moderation decisions of social media platforms in the EU. For this purpose, we gathered 156 million SoRs submitted by social media platforms within an observation period of two months. Our empirical analysis yields the following main findings: (i) There are vast differences in the frequency of content moderation across platforms. For instance, TikTok performs more than 350 times more content moderation decisions per user than X/Twitter. (ii) Content moderation is most commonly applied for text and videos, whereas images and other content formats undergo moderation less frequently. (ii) The primary reasons for moderation include content falling outside the platform's scope of service, illegal/harmful speech, and pornography/sexualized content, with moderation of misinformation being relatively uncommon. (iii) The majority of rule-breaking content is detected and decided upon via automated means rather than manual intervention. However, X/Twitter reports that it relies solely on non-automated methods. (iv) There is significant variation in the  content moderation actions taken across platforms. While most platforms commonly remove rule-breaking posts, others are more likely to opt for reducing its visibility. Altogether, our study implies inconsistencies in how social media platforms implement their obligations under the DSA -- resulting in a fragmented outcome that the DSA is meant to avoid. Our findings have important implications for regulators to clarify existing guidelines or lay out more specific rules that ensure common standards on how social media providers handle rule-breaking content on their platforms.
\end{abstract}


%



\maketitle

\section{Introduction}


\noindent 
Social media platforms (\eg, Facebook, X, YouTube) have become primary gateways to information and other content in the digital age \cite{Bakshy.2015}. While these platforms offer various social and economic benefits \cite{Utz.2016,Nisar.2019}, they also facilitate the dissemination of harmful (or even illegal) content, including, but not limited to, hate speech \cite{Solovev.2022a,Solovev.2023,Mathew.2019}, disinformation \cite{Feuerriegel.2023,Bennett.2020,Geissler.2023}, and calls for violence \cite{Bar.2023,Jakubik.2023,Baer.2022}. Concerns about harmful content on social media have been rising in recent years, particularly given its potential impact on elections \cite{Allcott.2017,Aral.2019,Bakshy.2015,Grinberg.2019,Guess.2020,Moore.2023}, public health \cite{Broniatowski.2018,Rocha.2021,Gallotti.2020,Roozenbeek.2020}, and public safety \cite{Muller.2021,Bar.2023,Oh.2013,Geissler.2023}. In response to such threats, many platforms have developed more or less sophisticated systems for content moderation, \ie, mechanisms that aim to prevent harm by removing or reducing the visibility of rule-breaking content \cite{Grimmelmann.2015}. However, their degree of strictness varies greatly from platform to platform, and each platform keeps the specifics of how it enacts its moderation decisions largely opaque \cite{Jhaver.2019}. Additionally, questions of who decides what is allowed on social media platforms and how providers decide to publish or remove third-party content have become pressing issues in public debate \cite{Goldman.2021}. Over the last years, there have been increasing calls for legislation to revise the present model of self‐regulation, where primarily social media platforms define the rules and procedures of online content moderation \cite{Schlag.2023,Turillazzi.2023}. 

As a remedy, the European Union (EU) and its member states have recently introduced and ratified an increasing number of laws and policies aimed at governing online content. The legislation should facilitate greater democratic control and oversight over systemic platforms, including social media \cite{Schlag.2023,Turillazzi.2023}. A key component of the European Commission's strategy is the \emph{Digital Services Act} (DSA) \cite{eu-2022-2065}, which has established a new set of obligations for social media providers to build a safer and more trustworthy digital space \cite{Schlag.2023,Cauffman.2021,Leerssen.2023,Turillazzi.2023}. Their primary focus is on strengthening the accountability of the platforms when it comes to illegal content uploaded by users. 
Specifically, major social media providers such as Facebook and X/Twitter can now be held responsible for the risks illegal content on their platforms poses to society. Furthermore, the DSA should establish a harmonized legal framework that avoids inconsistencies and uncoordinated procedures in content moderation amongst platforms. 
For this purpose, providers of large social media platforms are required to file detailed \emph{Statements of Reasons} (SoRs) explaining why content was moderated, by reference to the specific legal provision infringed. To ensure scrutiny of content moderation decisions and transparency for both platforms and users, all SoRs are made publicly available by the EU via the DSA Transparency Database \cite{EU.2023}.


\textbf{Research goal: } In this work, we provide a holistic early look at the DSA Transparency database. Due to this unique data source, we are, for the first time, able to empirically analyze real-world content moderation decisions of major social media platforms in the EU. Specifically, we address the following research questions: 
\begin{itemize}
	\item \textbf{(RQ1)} \emph{How frequently is social media content subject to content moderation in the EU?}
	\item \textbf{(RQ2)} \emph{How often are different types of content (text, images, video, etc.) moderated on social media?}
	\item \textbf{(RQ3)} \emph{What are specific reasons (\ie, legal grounds) due to which social media providers moderate content?} 
	\item \textbf{(RQ4)} \emph{To what extent are content moderation decisions on social media automated?}
	\item \textbf{(RQ5)} \emph{What types of content moderation actions do social media platforms implement?}
\end{itemize}

\textbf{Data \& Methods:} To address our research questions, we collected \emph{all} Statements of Reasons (SoRs) that were submitted by major social media platforms within the first two months of the DSA Transparency database in September 25, 2023, from the EU's official website. During our observation period, more than 156 million SoRs were transmitted by social media platforms, each representing one content moderation action on one of the platforms. We then extract a wide variety of variables (\eg, content types, legal grounds, etc.) from the SoRs in order to empirically analyze content moderation decisions on social media in the EU. Additionally, 
we implement regression analysis to estimate how the likelihood of automation of content moderation decisions varies across different content types and legal grounds. This allows us to characterize content moderation decisions that are more likely to be performed without human intervention by social media providers.

\textbf{Contributions:} To the best of our knowledge, this study is the first to present an empirical analysis of the Statements of Reasons in the EU's DSA Transparency Database. Our empirical analysis yields the following main findings: (i) There are vast differences in the frequency of content moderation across platforms. For instance, TikTok performs more than 350 times more content moderation decisions per user than X/Twitter. (ii) Content moderation is most commonly applied for text and videos, whereas images and other content formats undergo moderation less frequently. (ii) The primary reasons for moderation include content falling outside the platform's scope of service, illegal/harmful speech, and pornography/sexualized content, with moderation of misinformation being relatively uncommon. (iii) The majority of rule-breaking content is detected and decided upon via automated means rather than manual intervention. However, X/Twitter reports that it relies solely on non-automated methods. (iv) There is significant variation in the  content moderation actions taken across platforms. While most platforms commonly remove rule-breaking posts, others are more likely to opt for reducing their visibility. Altogether, our study implies inconsistencies in how social media platforms implement their obligations under the DSA -- resulting in a fragmented outcome that the DSA is meant to avoid. Our findings have important implications for regulators, who might be inclined to lay out more specific rules that ensure common standards on how social media providers handle rule-breaking content on their platforms.

\section{Background}

\subsection{Content Moderation on Social Media}

Content moderation describes mechanisms that are designed to prevent the dissemination of illegal and undesirable content in online communities \citep{Grimmelmann.2015,Roberts.2020}. In the context of social media, providers can choose from a wide catalogue of possible measures to prevent harm resulting from rule-breaking content on their platforms, such as, for example, content removal, visibility reduction (demotion), labeling, or account suspensions/terminations \citep{Jiang.2023}. Effective content moderation mechanisms are often considered to be essential to the functioning of social networks \cite{Gillespie.2018}. For instance, moderation interventions may increase compliance with community guidelines \cite{Ribeiro.2023} and reduce uncivil behaviour \cite{Katsaros.2022}. However, recent studies also indicate that content moderation efforts can backfire and increase rule breaking (\eg, because sanctioned users perceive the decision as unfair \cite{Chang.2019}), or lead to increased production of harmful content on other platforms \cite{Mitts.2022,Ali.2021,Russo.2023}.
 
Over the last couple of years, content moderation systems on mainstream platforms have become increasingly sophisticated. Historically, the moderation of content on social media was overseen by relatively small review teams and platform rules were limited in their scope \cite{Klonick.2017}. However, over time, as the demand from the public to address and remove harmful content grew, these practices evolved and platforms have established increasingly sophisticated systems to aid their content moderation efforts \cite{Meta.2023,Youtube.2023,X.2023,Tiktok.2023}. Many social media providers now enforce platform guidelines using automated content moderation systems that detect and intervene when rule-breaking occurs. Several platforms employ automated filters aimed at removing blatantly rule-breaking content (\eg, child sexual abuse material) \cite{Gillespie.2018,Gillespie.2020}. Moreover, platforms deploy moderators, who are either compensated  \cite{Roberts.2020} or volunteer \cite{Matias.2016}, to regulate the content that remains after automated filtering. Notably, each social media platform has developed various systems to implement these processes \cite{Gillespie.2018}, yet each platform keeps the specifics of how it enacts its moderation decisions opaque \cite{Jhaver.2019}.

\subsection{Digital Services Act}


The Digital Services Act (DSA) represents a significant legislative framework developed by the European Union (EU) with the primary goal of modernizing the digital space, ensuring safer and more open digital platforms for all users \cite{Schlag.2023,Cauffman.2021,Leerssen.2023,Turillazzi.2023}. At its core, the DSA aims to address the challenges brought by the rapid evolution and influence of digital services, particularly large online platforms. 
The DSA was adopted by the European Parliament in July 2022 and entered into force on November 16, 2022 \citep{eu-2022-2065}. It includes various requirements that providers of online platforms must fulfill. Each provider had to report their user numbers by February 17, 2023. These figures were used to identify all Very Large Online Platforms (VLOPs) and Search Engines (VLOEs) with more than 45 million users in the EU (corresponding to 10\% of the population). Starting on August 25, 2023, all VLOEs and VLOPs must fulfill all obligations contained in the DSA. For all other (smaller) platforms, the rules apply from February 17, 2024. This includes publishing a transparency report every six months in which platforms must describe their own content moderation measures, as well as other relevant information such as, for example, the number of reports they receive from users, the error rate of automated content moderation systems and the composition of their content moderation teams (qualifications and linguistic expertise). 


Furthermore, as mandated by Article 17 of the DSA, all VLOPs and VLOEs are required to provide detailed Statements of Reasons (SoRs) for any content moderation activity, including content removal, reach restriction, or account suspension/termination. The intention is to inform users about content moderation decisions and explain the reasons for the respective decisions. In accordance with Article 24(5) of the DSA, all submitted statements are collected and made publicly available on the DSA Transparency Database, which is managed by the Directorate-General for Communications Networks, Content and Technology of the European Commission. SoRs shall be clear and specific and easily comprehensible and as precise and specific as reasonably possible under the given circumstances. To ensure this, they must contain information about the content and the implemented type of moderation, whether automated systems were used, and whether the subject contains illegal content or is incompatible with the terms and conditions of the provider. 

\section{Data}

\subsection{Data Source: Statements of Reasons}

We downloaded \emph{all} Statements of Reasons (SoRs) that were submitted between the introduction of the database on September 25, 2023, and November 25, 2023, from the website of the DSA Transparency Database, \ie within an observation period of two months. In total, more than 550 million SoRs were transmitted during the observation period. 
SoRs are submitted by all VLOPs and VLOEs. As we focus on content moderation on social media, we only included SoRs submitted by large social media platforms, namely \emph{Facebook, Instagram, YouTube, TikTok, Snapchat, X, LinkedIn} and \emph{Pinterest}. Furthermore, we excluded SoRs for content published before August 25, 2023, as the obligations associated with the DSA framework became effective from this date. The resulting dataset contains more than 156 Mio. SoRs, each including information on one content moderation action.

\subsection{Key Variables}

We now present the key variables that we extracted from the SoRs provided in the DSA Transparency Database. All of these variables will be empirically analyzed in the next sections: 

\begin{itemize}[leftmargin=*]
	\item \textsc{Platform Name:} The platform that submitted the SoR (\emph{Facebook, Instagram, Youtube, TikTok, Snapchat, X,} or \emph{Pinterest}).
	\item \textsc{Content Type:} The type of content that is moderated (\emph{Audio, Product, Synthetic Media, Image, Video, Text,} or \emph{Other}).
	\item \textsc{Category:} A categorical variable to specify the type of illegality/incompatibility because of which the content/account was moderated (\emph{Violence, Unsafe \& Illegal Products, Self Harm, Scope of Platform Service, Scams \& Fraud, Risk for Public Security, Protection of Minors, Pornography/Sexualized Content, Non Consensual Behavior, Mis-/Disinformation, Intellectual Property Infringements, Illegal/ Harmful Speech, Data Protection/Privacy Violations} or \emph{Animal Welfare}). 
	\item \textsc{Decision Ground:} Whether the content is classified as \emph{Illegal} or \emph{Incompatible}.
	\item \textsc{Automated Decision:} Whether the content moderation decision was performed \emph{Not Automated}, \emph{Partially Automated}, or \emph{Fully Automated}.
	\item \textsc{Automated Detection:} Whether automated means were used to identify the content addressed by the decision (\emph{Yes} or \emph{No}).
	\item \textsc{Decision Type:} The content moderation action implemented by the platform (\emph{Account Suspended, Account Terminated, Content Age Restricted, Content Demoted, Content Disabled, Content Labelled, Content Removed, Content Demoted/Removed} or \emph{Other}).
\end{itemize}

\section{Empirical Analysis}

\subsection{Frequency of Content Moderation (RQ1)}

\begin{figure}[H]
	\captionsetup{position=top}
	\captionsetup{belowskip=1pt}
	\captionsetup[subfloat]{textfont={sf,normalsize}, skip=2pt, singlelinecheck=false, labelformat=simple,labelfont=bf,justification=centering}
	\centering
	\includegraphics[width=.95\linewidth]{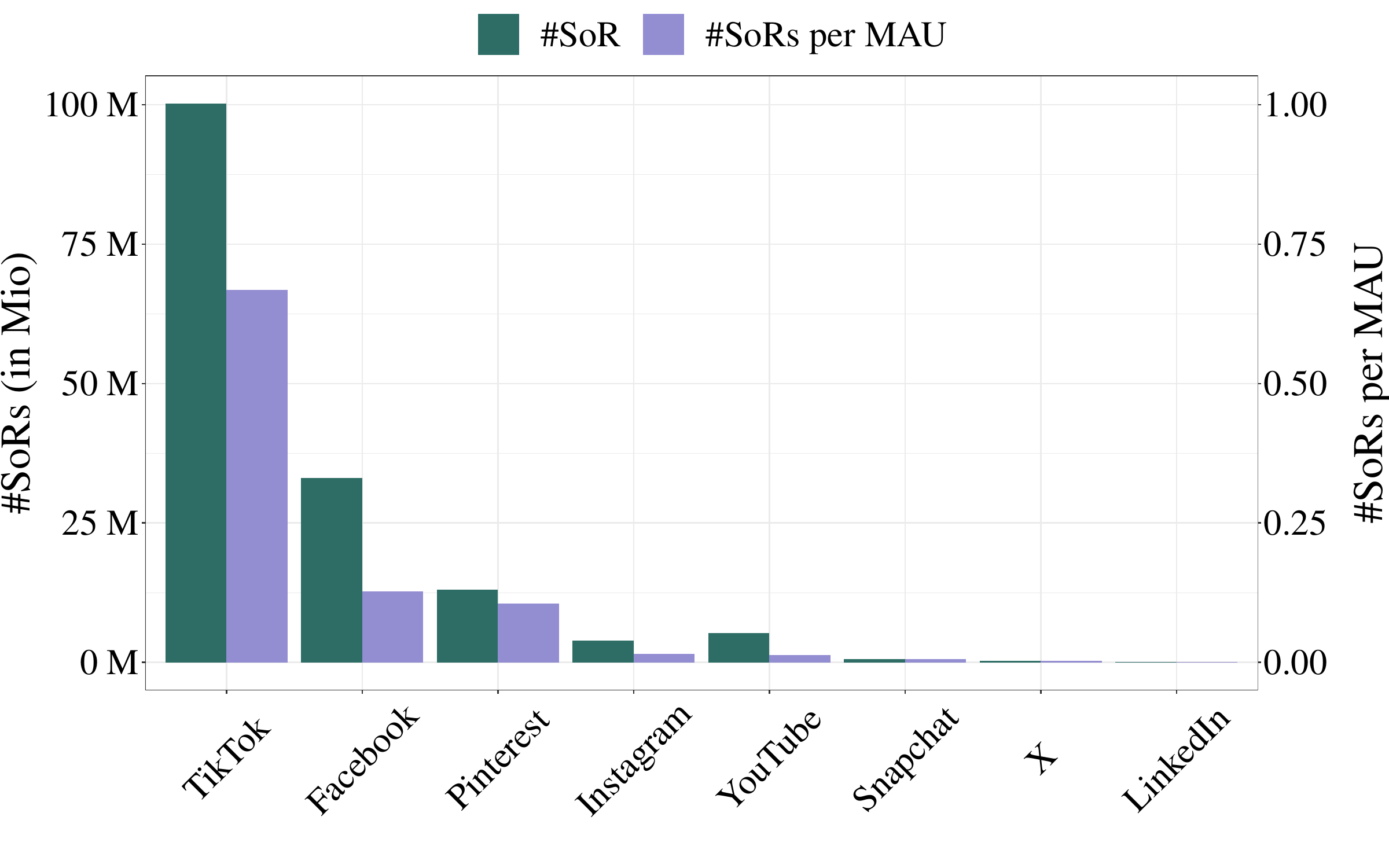}
	\caption{Number of SoRs per platform relative (left scale) and number of SoRs per platform to the number of monthly active users (MAU) (right scale).}
	\label{fig:appl_per_platform}
	\vspace{-.5cm}
\end{figure}

We start by analyzing how many content moderation actions were submitted by each social media platform within the EU  (see Fig. \ref{fig:appl_per_platform}). The largest number of SoRs (\textbf{\textcolor{Teal}{\#SoR}}) was submitted by TikTok (100.15 Mio; 64.09\%), followed by Facebook (33.70 Mio; 21.56\%), Pinterest (12.45 Mio; 7.97\%), YouTube (5.12 Mio; 3.28\%), and Instagram (3.95 Mio; 2.53\%). Snapchat (0.61 Mio; 0.39\%), X/Twitter (0.27 Mio; 0.17\%), and LinkedIn (0.03 Mio; 0.02\%) submitted less than one million SoRs during our observation period. It is striking that TikTok moderates substantially more content than all other platforms, some of which are much more relevant in the European market in terms of user numbers (\eg YouTube, Facebook, and Instagram).

Fig. \ref{fig:appl_per_platform} shows how many content moderation decisions were made by platforms in relation to their size. For this, we calculated the ratio of content moderation actions (SoRs) relative to the monthly active users (MAU) per platform within the EU (\textbf{\textcolor{LightPurple}{\#SoRs per MAU}}).\footnote{We collected information on the number of monthly active users (MAU)  within the EU from the platforms' Transparency Reports \citep{TikTok.2023b,Meta.2023b,Meta.2023c,LinkedIn.2023,Snapchat.2023,Youtube.2023,X.2023b}. According to these numbers, YouTube is the largest platform with more than 400 million MAUs, for Instagram and Facebook Meta reported 259 million MAUs each, TikTok has 150 million MAUs, Pinterest around 150 million MAUs, X/Twitter around 100 million, Snapchat just under 97 million MAUs, and LinkedIn reported around 45 million MAUs.} It is evident that TikTok is by far the most active in terms of content moderation, even when considering differences in platform sizes. TikTok submitted over 0.67 SoRs per MAU, surpassing Facebook (0.13) and Pinterest (0.10). The remaining platforms had comparatively few SoRs per MAU (each less than 0.02) For example, the rate at which TikTok carried out content moderation decisions per MAU was more than 350 times that of X/Twitter.

\subsection{Content Types (RQ2)}

\begin{figure}[H]
	\captionsetup{position=top}
	\captionsetup{belowskip=1pt}
	\captionsetup[subfloat]{textfont={sf,normalsize}, skip=2pt, singlelinecheck=false, labelformat=simple,labelfont=bf,justification=centering}
	\centering
	\includegraphics[width=.95\linewidth]{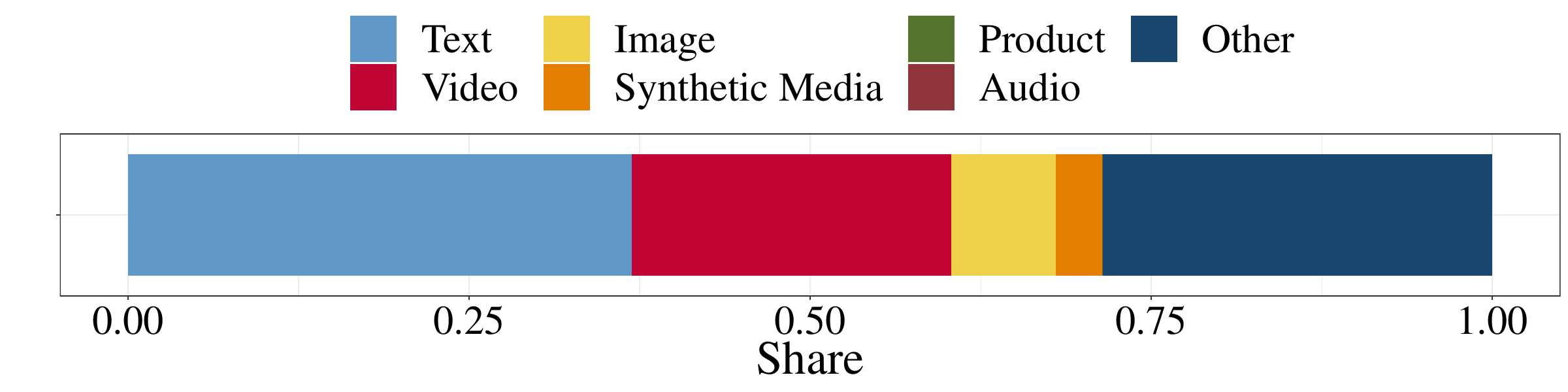}
	\caption{Relative share of SoRs per content type.}
	\label{fig:content_type}
	\vspace{-.5cm}
\end{figure}

Next, we study how often different types of content (\ie, text, images, videos, etc.) were moderated on social media platforms in the EU. To assign the SoRs to individual content types, we used the categories specified by the platforms when submitting the SoRs (\emph{Content Type}). As shown in Fig. \ref{fig:content_type}, the most frequently moderated content is \textbf{\textcolor{LightBlue}{Text}} (38.74\%), followed by \textbf{\textcolor{RichRed}{Videos}} (24.42\%), \textbf{\textcolor{BrightYellow}{Images}} (6.57\%), and \textbf{\textcolor{Orange}{Synthetic Media}} (38.74\%). Approximately 27\% of all content moderation actions were assigned to the category \textbf{\textcolor{DeepBlue}{Other}}, \ie, content types that are not predefined by the DSA. The majority of this content concerns violations that affect entire accounts/profiles, platform-specific content types (\eg pins or boards on Pinterest), and (job) advertisements (in particular on Youtube, Pinterest, and LinkedIn). 

\begin{figure}[H]
	\captionsetup{position=top}
	\captionsetup{belowskip=1pt}
	\captionsetup[subfloat]{textfont={sf,normalsize}, skip=2pt, singlelinecheck=false, labelformat=simple,labelfont=bf,justification=centering}
	\centering
	\includegraphics[width=.95\linewidth]{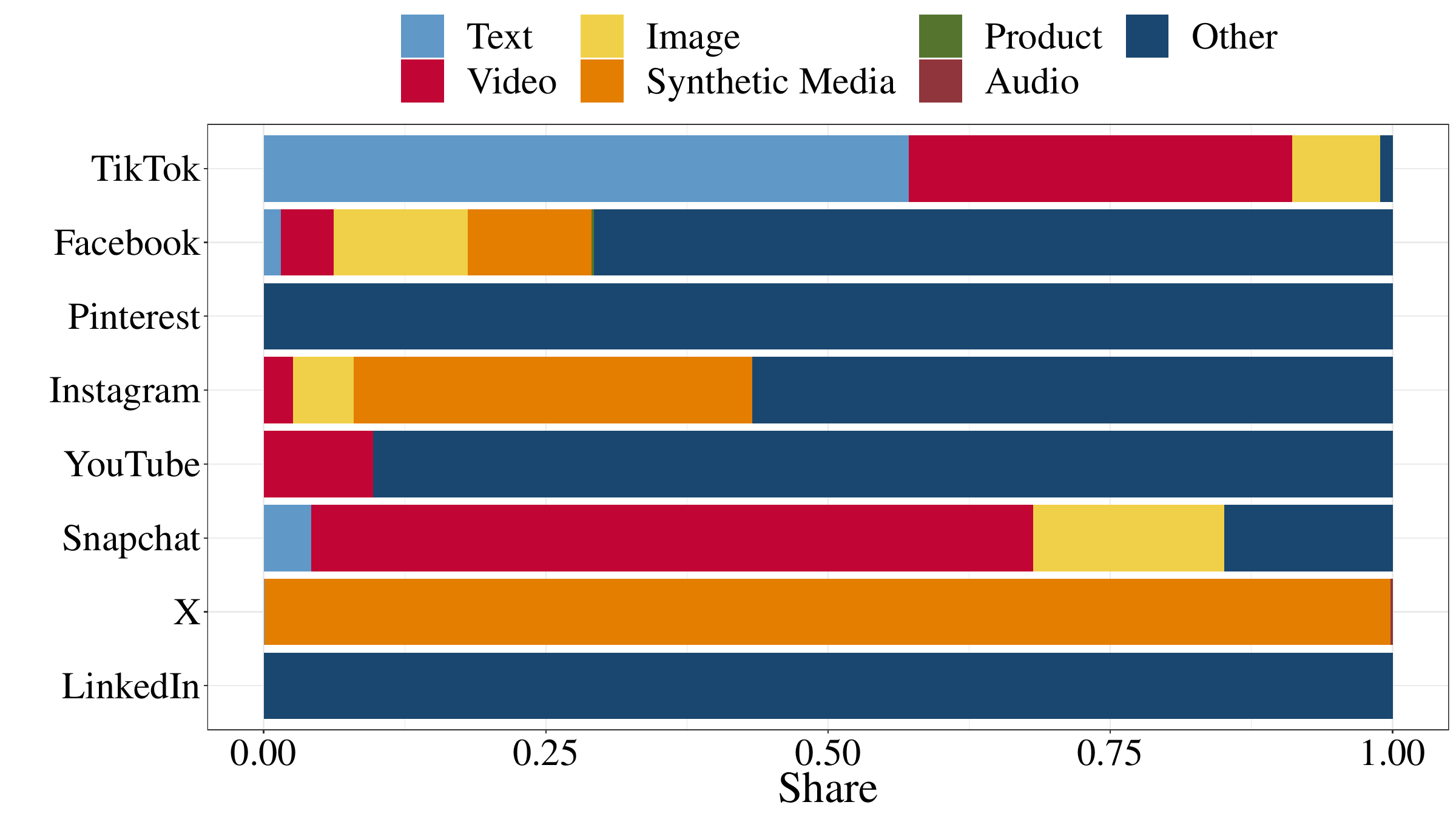}
	\caption{Relative share of SoRs per content type across platforms.}
	\label{fig:content_type_platforms}
	\vspace{-.5cm}
\end{figure}

To analyze whether the distribution of moderated content types is platform-specific, Fig. \ref{fig:content_type_platforms} visualizes the distribution of the seven predefined content categories per platform. We observe that TikTok primarily focuses on \textbf{\textcolor{LightBlue}{Text}} (57.14\%) and \textbf{\textcolor{RichRed}{Videos}} (33.97\%), whereas moderation of \textbf{\textcolor{BrightYellow}{Images}} (7.78\%) is relatively rare. In contrast, Snapchat rarely moderates \textbf{\textcolor{LightBlue}{Text}} (4.21\%) but is relatively more likely to moderate \textbf{\textcolor{RichRed}{Videos}} (63.92\%) and \textbf{\textcolor{BrightYellow}{Images}} (16.96\%). The platform X/Twitter seems to limit its content moderation activities mainly to content categorized as \textbf{\textcolor{Orange}{Synthetic Media}} (99.83\%). Note, however, that content assigned to this category may include artificially created content in the form of, for example, text, images, video and/or audio content. The remaining platforms each categorized more than 50\% of all content moderation actions as content type \textbf{\textcolor{DeepBlue}{Other}}. 

Altogether, we find that the type of content subject to moderation is, in many cases, strongly related to the content that is predominantly published on the respective platform (\eg, Video and Image on Snapchat, Video and Advertisement on YouTube, Pins on Pinterest, job-related content on LinkedIn). It is also worth noting that the content type is specified by the platform (\ie, self-reported), and each content piece is assigned to a single type. 
However, on social media, posts can consist of a mixture of different content types (\eg, video/image and text). In a similar vein, AI-generated content may be classified as image/text/video or as synthetic media. Overall, it seems likely that the content type reported by platforms often describes only one dimension of a social media post. 

\subsection{Reasons for Moderation (RQ3)}

We now analyze specific reasons (\ie, legal grounds) due to which social media providers moderate content. When submitting a SoR, the platforms must assign the content to one of 14 given categories to specify the type of illegality/incompatibility because of which the content/account was moderated. Assigning a second category is possible, but not frequently used by the platforms (less than 0.001\%). 
Overall, the most frequent reasons for moderation is content falling outside of the \textbf{\textcolor{LightPurple}{Scope of Platform Service}} (49.06\%), \textbf{\textcolor{LightBrown}{Illegal/Harmful Speech}} (28.50\%), \textbf{\textcolor{DarkYellow}{Pornography/Sexualized Content}} (8.77\%), and \textbf{\textcolor{Teal}{Violence}} (6.70\%). Content that is classified in the remaining categories (\textbf{\textcolor{SoftBlue}{Data Protection/Privacy Violations}}, \textbf{\textcolor{SoftCerulean}{Protection of Minors}}, \textbf{\textcolor{Coral}{Scams \& Fraud}}, \textbf{\textcolor{LightBlue}{Intellectual Property Infringements}}, \textbf{\textcolor{RichRed}{Self Harm}}, \textbf{\textcolor{BrightYellow}{Mis-/Disinformation}}, \textbf{\textcolor{Orange}{Unsafe \& Illegal Products}}, \textbf{\textcolor{DarkGreen}{Non Consensual Behavior}}, \textbf{\textcolor{DeepRed}{Animal Welfare}}, and \textbf{\textcolor{DeepBlue}{Risk for Public Security}}) is only relatively rarely subject to moderation (6.97\% in total). 

\begin{figure}[H]
	\captionsetup{position=top}
	\captionsetup{belowskip=1pt}
	\captionsetup[subfloat]{textfont={sf,normalsize}, skip=2pt, singlelinecheck=false, labelformat=simple,labelfont=bf,justification=centering}
	\centering
	\includegraphics[width=.95\linewidth]{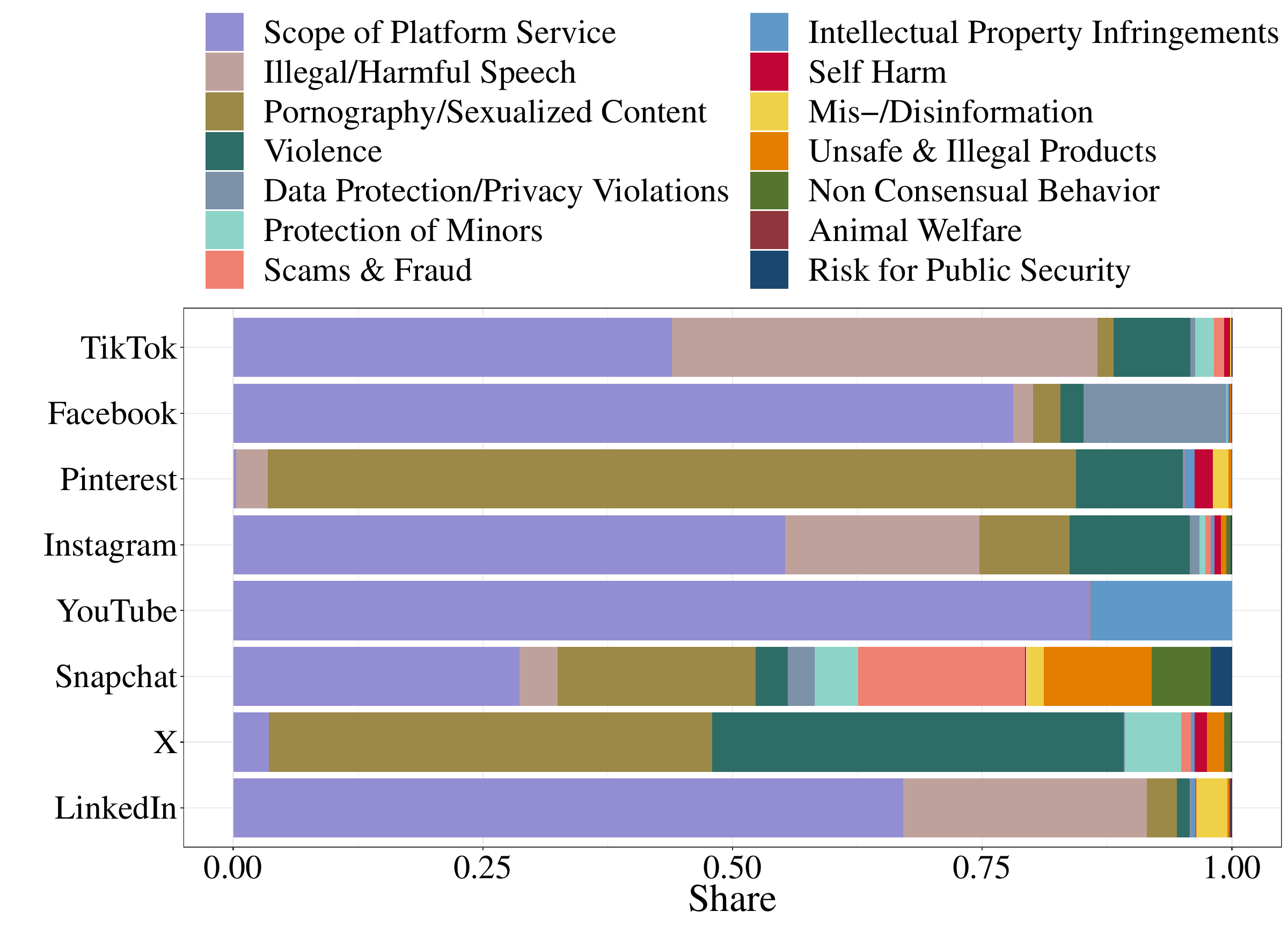}
	\caption{Relative share of different types of incompatibility/illegality across platforms.}
	\label{fig:category_decision_ground}
	\vspace{-.5cm}
\end{figure}

Fig. \ref{fig:category_decision_ground} illustrates the distribution of the different categories per platform. We find that there is a relatively highly similarity across most platforms. With the exceptions of X/Twitter and Pinterest, all platforms moderated a large proportion of content that does not correspond to the \textbf{\textcolor{LightPurple}{Scope of Platform Service}}, \textbf{\textcolor{LightBrown}{Illegal/Harmful Speech}}, and \textbf{\textcolor{Teal}{Violence}}. In contrast, Pinterest focused primarily on \textbf{\textcolor{DarkYellow}{Pornography/Sexualized Content}} (80.95\%). In the case of X/Twitter, the vast majority of content moderation actions were attributed to \textbf{\textcolor{Teal}{Violence}} (41.18\%) or \textbf{\textcolor{DarkYellow}{Pornography/Sexualized Content}} (44.37\%). Snapchat is the platform that moderated content in the widest variety of categories (\eg 16.69\% in \textbf{\textcolor{Coral}{Scams \& Fraud}} and 10.77\%  \textbf{\textcolor{Orange}{Unsafe \& Illegal Products}}).

\begin{figure}[H]
	\captionsetup{position=top}
	\captionsetup{belowskip=1pt}
	\captionsetup[subfloat]{textfont={sf,normalsize}, skip=2pt, singlelinecheck=false, labelformat=simple,labelfont=bf,justification=centering}
	\centering
	\includegraphics[width=.95\linewidth]{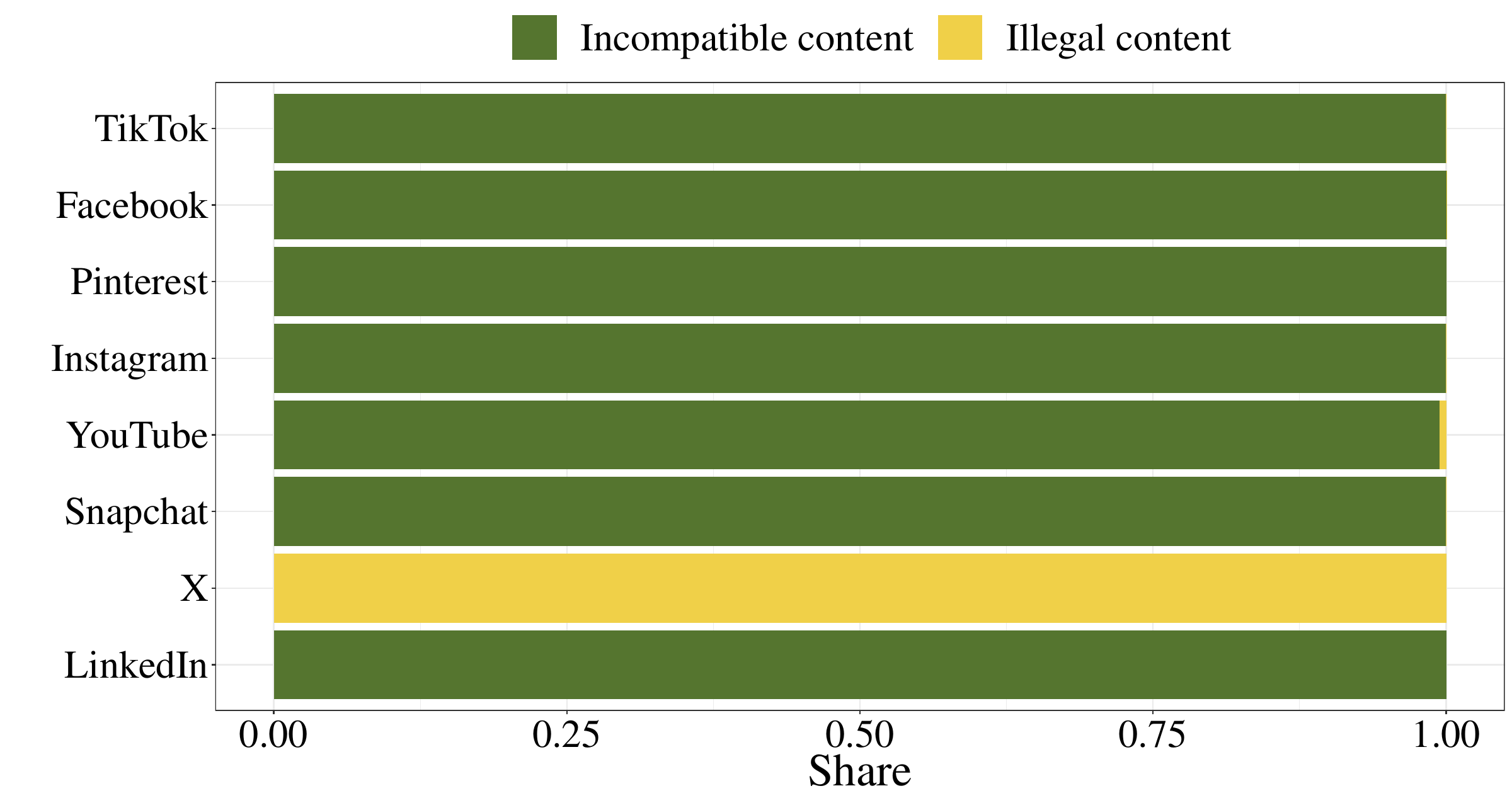}
	\caption{Relative share of content moderation decisions due to \emph{illegal} vs. \emph{incompatible} content.}
	\label{fig:decision_ground_platform}
	\vspace{-.5cm}
\end{figure}

In addition to the category, platforms must state whether the moderation decision was taken in line with article 17(3)(d) DSA, meaning the content is considered as \textbf{\textcolor{BrightYellow}{illegal}}, or in line with article 17(3)(e), meaning the content is considered \textbf{\textcolor{DarkGreen}{incompatible}} with the service's terms and conditions. Across all platforms, most content moderation actions were performed because of content being \textbf{\textcolor{DarkGreen}{incompatible}} (99.80\%) rather than \textbf{\textcolor{BrightYellow}{illegal}} (0.20\%).  
When comparing the individual platforms, however, it is striking that all platforms except X/Twitter moderated more than 99\% of \textbf{\textcolor{DarkGreen}{incompatible}} content. In contrast, X/Twitter reported that 100\% of their content moderation actions were taken due to \textbf{\textcolor{BrightYellow}{illegal}} content (see Fig. \ref{fig:decision_ground_platform}). We thus again notice a significant difference in how X/Twitter vs. other platforms interpret their obligations under the DSA. 

\subsection{Automation of Content Moderation (RQ4)}

\begin{figure}[H]
	\captionsetup{position=top}
	\captionsetup{belowskip=1pt}
	\captionsetup[subfloat]{textfont={sf,normalsize}, skip=2pt, singlelinecheck=false, labelformat=simple,labelfont=bf,justification=centering}
	\centering
	\includegraphics[width=.99\linewidth]{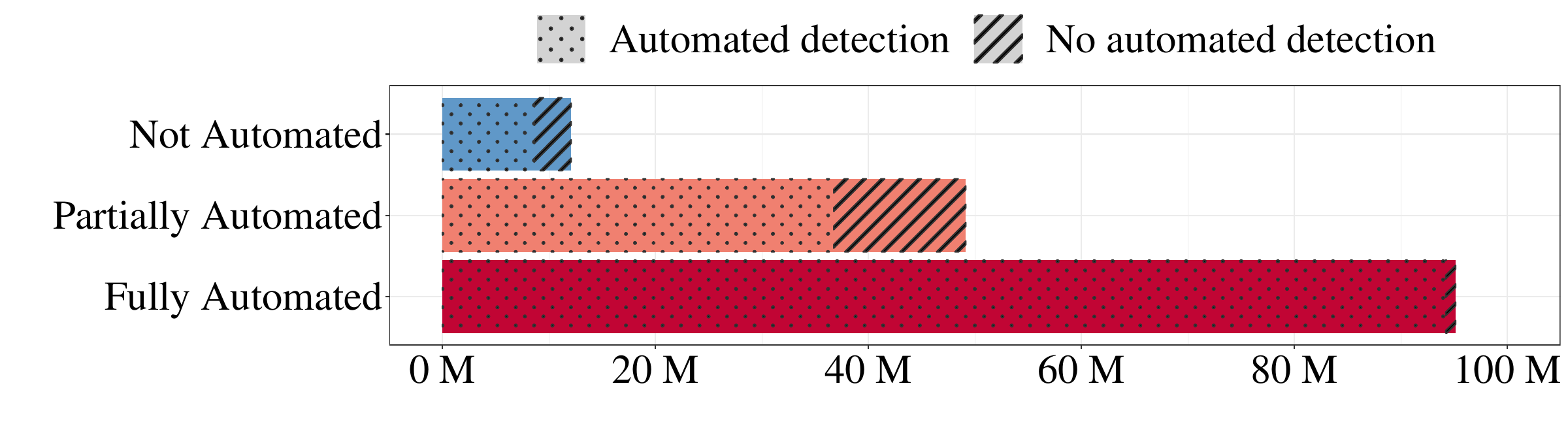}
	\caption{Number of SoRs per level of decision automation separated by whether the content has been identified automatically (dotted) or manually (striped).}
	\label{fig:decision_detection_automated}
	\vspace{-.5cm}
\end{figure}

Platforms are also asked to indicate whether the content moderation decision was taken automatically (\emph{Automated Decision}), and whether the decision was taken on content that has been detected or identified using automated means (\emph{Automated Detection}). Fig. \ref{fig:decision_detection_automated} shows that more than 90 million (60.67\%) of all submitted content moderation decisions were performed \textbf{\textcolor{RichRed}{fully automated}} (\ie, without human intervention), 48 million (31.48\%) \textbf{\textcolor{Coral}{partially automated}}, and 11.8 million (7.74\%) \textbf{\textcolor{LightBlue}{not automated}}. 
Furthermore, the vast majority of violations have been detected using some sort of automation (89.45\%). 

Interestingly, there is a strong link between the methods used for identification and decision making.  Within the \textbf{\textcolor{RichRed}{fully automated}} content moderation activities, more than 99\% of all content was also first identified automatically. Within the other decision-making categories, this proportion is significantly lower at 76.06\% (\textbf{\textcolor{Coral}{partially automated}}), and 71.75\% (\textbf{\textcolor{LightBlue}{not automated}}), respectively. In other words, automated identification was usually followed by automated decision-making, while a larger proportion of content that was moderated using manual intervention was also previously identified or reported by humans (\eg, users or content moderators).

\begin{figure}[H]
	\captionsetup{position=top}
	\captionsetup{belowskip=1pt}
	\captionsetup[subfloat]{textfont={sf,normalsize}, skip=2pt, singlelinecheck=false, labelformat=simple,labelfont=bf,justification=centering}
	\centering
	\includegraphics[width=.95\linewidth]{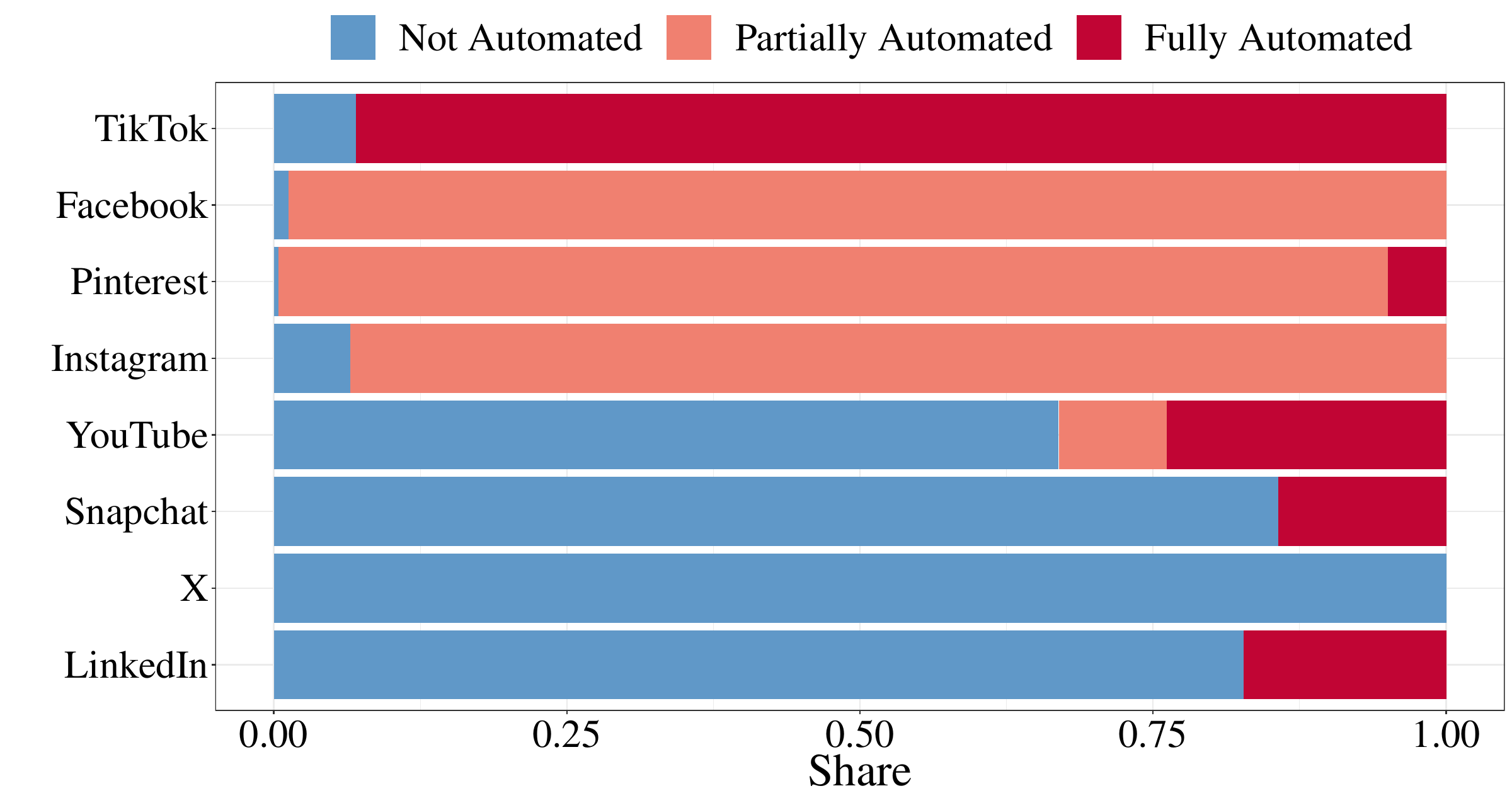}
	\caption{Relative share of the level of automated decision making per platform.}
	\label{fig:automated_decision_platform}
	\vspace{-.5cm}
\end{figure}

Analyzing automated decision-making across social media platforms reveals further differences (see Fig. \ref{fig:automated_decision_platform}). The vast majority of TikTok's content moderation actions was performed \textbf{\textcolor{RichRed}{fully automated}}, while Facebook, Pinterest and Instagram tend to combine automated means and human intervention (\ie, \textbf{\textcolor{Coral}{partially automated}}). In contrast, YouTube, Snapchat, X/Twitter and LinkedIn performed the majority of their content moderation \textbf{\textcolor{LightBlue}{not automated}}. It is striking that X/Twitter again stands out with a completely manual content moderation (\ie, not automated).

In order to better understand situations in which content moderation is more likely to be performed automatically, we implement a logistic regression model estimating how the likelihood of automated content moderation decision varies across different content types and decision grounds. The dependent variable is \emph{Automated Decision}, a binary variable ($=1$ if yes; otherwise $=0$) that describes whether a moderation decision $i$ was performed fully automated or not (\ie, not/partially automated vs. fully automated). The key explanatory variables are $\mathit{Automated Detection}_i$ (reference type: \emph{No Automated Detection}), $\mathit{Content Type}_i$ (reference type: \emph{Other})  and $\mathit{Category}_i$ (reference type: \emph{Scope of Platform Service}). Additionally, we control for the time elapsed (in days) between the publication of the content on the platform and the application of the content moderation $\mathit{Delay}_i$ ($z$-standardized). The resulting model is
{
	\begin{align}
		\logit(&\mathit{Automated Decision_i})  = \,\beta_{0}  +  \beta_{1} \, \mathit{Automated Detection}_i \label{eq:neg_bin} \\
		          &  + \beta_{2} \, \mathit{Content Type}_i + \beta_{3} \,  \mathit{Category}_i  + \beta_{3} \,  \mathit{Delay}_i  + \lambda_{i} + u_{i}, \nonumber 
	\end{align} 
}%

with intercept $\beta_0$, monthly fixed effects $u_{i}$ to adjust for differences in the date the content was published, and platform fixed effects $\lambda_{i}$. Note that since we apply a logistic regression, the odds ratio (\ie, the exponentiated coefficients) must be calculated to determine the effect sizes (see, eg, \cite{Ai.2003}). 

\begin{figure}[H]
	\captionsetup{position=top}
	\captionsetup{belowskip=1pt}
	\captionsetup[subfloat]{textfont={sf,normalsize}, skip=2pt, singlelinecheck=false, labelformat=simple,labelfont=bf,justification=centering}
	\centering
	\includegraphics[width=.95\linewidth]{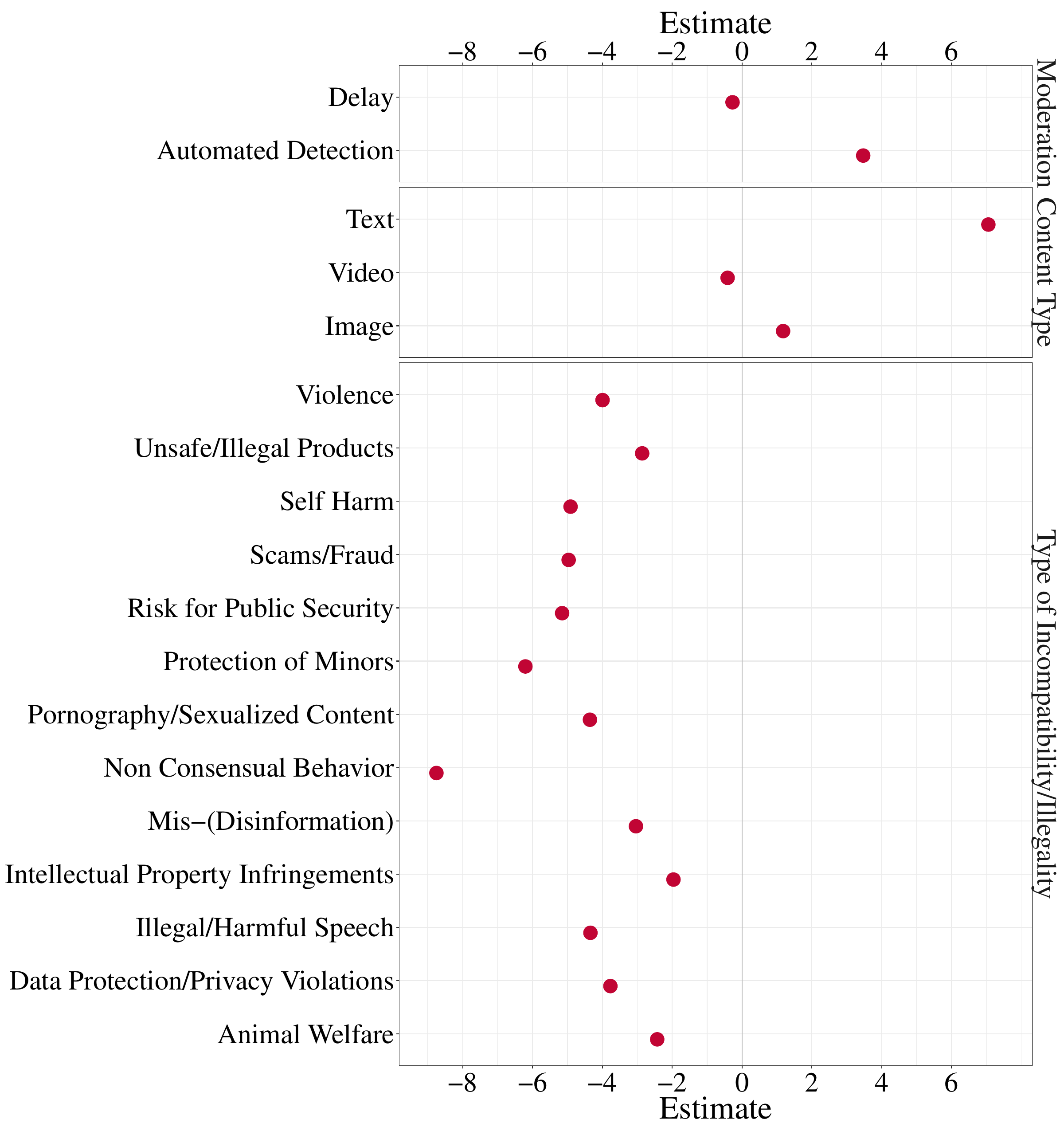}
	\caption{Coefficient estimates (circles) and 99\% confidence intervals (bars) for a logistic regression model with automated decision ($\boldsymbol{=1}$ if yes; otherwise $\boldsymbol{=0}$) as the dependent variable ($\boldsymbol{n =}$ \num{156378199}). Monthly fixed effects and platform fixed effects are included. The reference values of the categorical variables are \emph{No Automated Detection} ($\boldsymbol{Automated Detection_i}$), \emph{Other} ($\boldsymbol{Content Type_i}$), and \emph{Scope of Platform Service} ($\boldsymbol{Category_i}$).}
	\label{fig:main_model}
	\vspace{-.5cm}
\end{figure}

The coefficient estimates and their 99\% confidence intervals are visualized in Fig. \ref{fig:main_model}. We find that the odds of a content moderation decision being performed automatically are $e^{3.470} \approx 32.143$ times higher if the content was detected automatically (OR: $32.143$, coef: $3.470$, $p<0.01$). Across the different content types, we find a strong positive association for \emph{Text} (OR: $1158.804$, coef: $7.055$, $p<0.01$). Specifically, the odds of a content moderation decision performed automatically are significantly higher for text content compared to other content (\ie, \textit{Product}, \textit{Synthetic Media}, \textit{Audio}, \textit{Other}). At the same time, we find a smaller positive association for \emph{Images} (OR: $3.243$, coef: $1.177$, $p<0.01$), while \emph{Videos} are less likely to be moderated automatically (OR: $0.662$, coef: $-0.413$,$p<0.01$). With regard to legal grounds, we find that all categories are statistically significantly less likely to be automatically moderated compared to content that does correspond to the \emph{Scope of Platform Service} (each $p<0.01$). The strongest negative association can be observed for the (relatively rare) type of \emph{Non Consensual Behaviour} (OR: $0.0001$, coef: $-8.753$, $p<0.01$). The effect sizes for the more frequently moderated types \emph{Illegal/Harmful Speech} (OR: $0.013$, coef: $-4.342$, $p<0.01$), \emph{Pornography/Sexualized Content} (OR: $0.012$, coef: $-4.357$, $p<0.01$) and \emph{Violence} (OR: $0.018$, coef: $-3.995$, $p<0.01$) are all very similar to each other. 
We also observe a small negative association between the time elapsed between the date the content was published on the platform and the moderation date (OR: $0.761$, coef: $-0.273$, $p<0.01$). 

\subsection{Content Moderation Actions (RQ5)}

\begin{figure}[H]
	\captionsetup{position=top}
	\captionsetup{belowskip=1pt}
	\captionsetup[subfloat]{textfont={sf,normalsize}, skip=2pt, singlelinecheck=false, labelformat=simple,labelfont=bf,justification=centering}
	\centering
	\includegraphics[width=.95\linewidth]{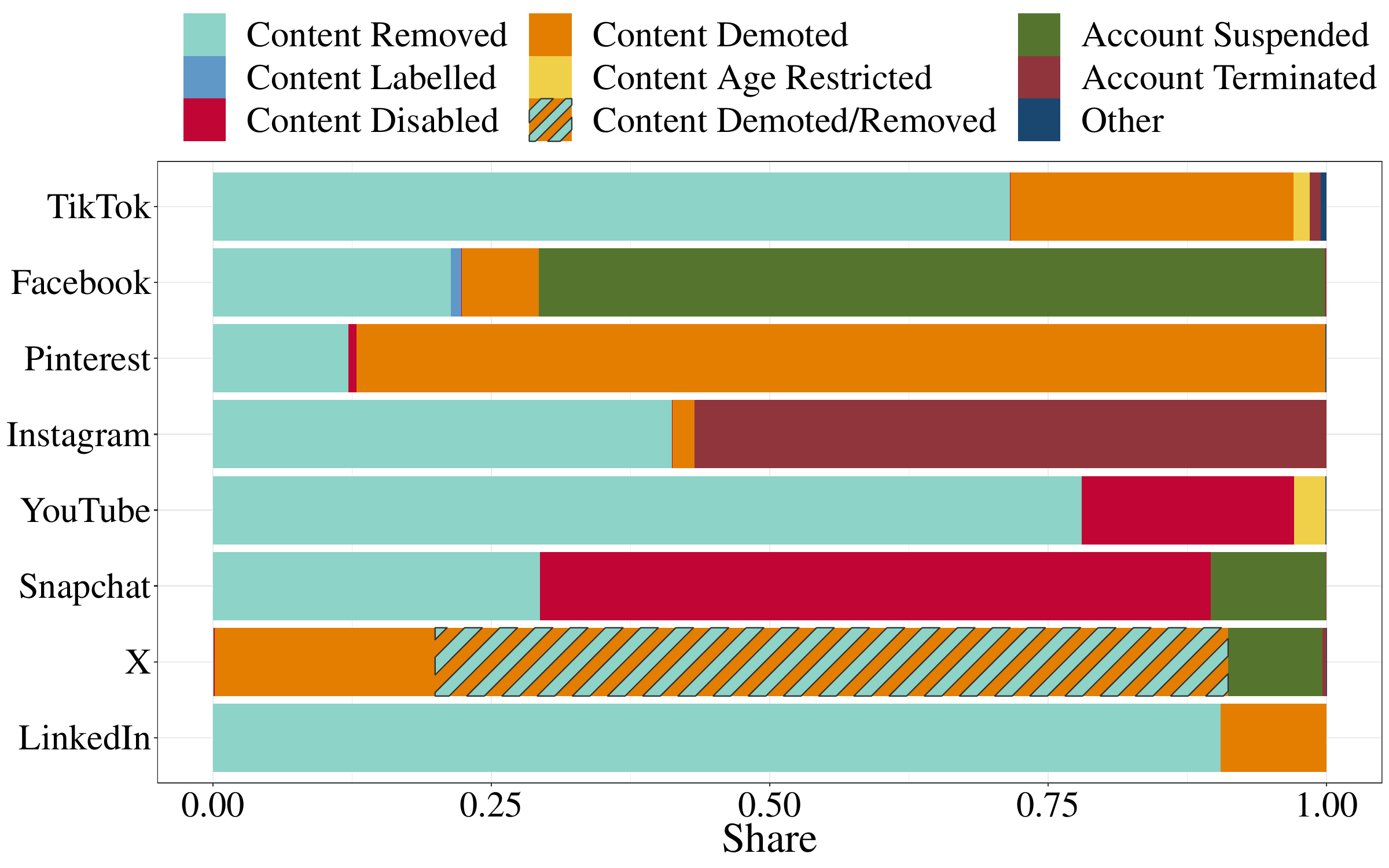}
	\caption{Relative share of different content moderation actions per platform. Note that X/Twitter describes a large part of the moderated content as ``not suitable for work'' (NSFW) (71.20\%) whereby, according to the platform's own guidelines \citep{X.2023c}, either a reduction in visibility (\ie, \textbf{\textcolor{Orange}{demotion}}) or \textbf{\textcolor{SoftCerulean}{removal}} of the content is applied (striped area).}
	\label{fig:decision_type_platform}
	\vspace{-.5cm}
\end{figure}

While the focus in the previous sections was on which and why content is moderated, we now take a closer look at the specific content moderation actions. Content moderation actions performed under the DSA can be grouped into two categories: (i) actions that affect a specific content piece (\ie, \textbf{\textcolor{SoftCerulean}{removal}}, \textbf{\textcolor{LightBlue}{labelling}}, \textbf{\textcolor{RichRed}{disabling}}, \textbf{\textcolor{Orange}{demotion}} and \textbf{\textcolor{BrightYellow}{age restriction}} of content); (ii) actions that affect an account (\ie, \textbf{\textcolor{DarkGreen}{suspension}} or \textbf{\textcolor{DeepRed}{termination}} of an account). The platforms describe in the SoRs how the content was moderated, either by selecting one of these predefined actions or by selecting \textbf{\textcolor{DeepBlue}{Other}}. In the case of the latter, platforms can to provide a short description of the action that was taken. However, in many cases, the content of the text descriptions is very close to the the predefined action categories (\eg, ``Limited distribution'' $\rightarrow$ \textbf{\textcolor{Orange}{Content Demoted}}). To accommodate such cases, we employed string matching to assign the text descriptions to the predefined action categories.\footnote{The descriptions were assigned to the predefined action categories as follows: ``Limited distribution'' $\rightarrow$ \textbf{\textcolor{Orange}{Content Demoted}}, ``not eligible for recommendation'' $\rightarrow$ \textbf{\textcolor{Orange}{Content Demoted}},  ``mute audio'' $\rightarrow$ \textbf{\textcolor{RichRed}{Content Disabled}}, ``Bounce'' $\rightarrow$ \textbf{\textcolor{Orange}{Content Demoted}}, ``Ban'' $\rightarrow$ \textbf{\textcolor{RichRed}{Content Disabled}}, ``AddTweetAnnotation'' $\rightarrow$ \textbf{\textcolor{LightBlue}{Content Labelled}} (ordered by descending frequency). X/Twitter describes a large part of the moderated content as ``not suitable for work'' (NSFW) (71.20\%) whereby, according to the platform's own guidelines, either a reduction in visibility (\ie, \textbf{\textcolor{Orange}{demotion}}) or \textbf{\textcolor{SoftCerulean}{removal}} of the content is implemented (see striped area in Fig. \ref{fig:decision_type_platform}) \citep{X.2023c}. 
} In cases where a description is absent or cannot be unambiguously assigned, the category \textbf{\textcolor{DeepBlue}{Other}} has been retained (0.36\%).

 Fig. \ref{fig:decision_type_platform} shows the distribution of implemented content moderation across platforms. Overall, we observe that the most frequent types of content moderation are the \textbf{\textcolor{SoftCerulean}{removal of content}} (55.15\%), the \textbf{\textcolor{Orange}{demotion of content}} (25.15\%), and the \textbf{\textcolor{DarkGreen}{suspension of accounts}} (14.96\%).  The tendency to focus on \textbf{\textcolor{SoftCerulean}{content removal}} and account \textbf{\textcolor{DarkGreen}{suspension}}/\textbf{\textcolor{DeepRed}{termination}} as the primary choice of content moderation is prevalent across most platforms. However, there are also differences. For instance, Pinterest primarily employed \textbf{\textcolor{Orange}{content demotion}}. Conversely, Snapchat implemented \textbf{\textcolor{RichRed}{content disabling}} in over 50\% of its content moderation actions. X/Twitter does not distinguish between \textbf{\textcolor{SoftCerulean}{removal}} and \textbf{\textcolor{Orange}{demotion}} in its reporting to the DSA Transparency Database. 

\section{Discussion}

\textbf{Relevance:} Effective mechanisms for content moderation are important tools to prevent the dissemination of illegal and undesirable content in social networks \cite{Gillespie.2018}. However, content moderation decisions by social media providers are oftentimes perceived as intransparent and publicly inaccessible. Previous work was mostly limited to investigating moderation interventions in artificial environments (\eg, interviews, surveys) \cite[\eg,][]{Mena.2020,Moravec.2020,Pennycook.2020,Bode.2015,Clayton.2020} or based on observational datasets restricted in scope \cite[\eg,][]{Zannettou.2021,Prollochs.2022,Prollochs.2023,Ling.2023,Drolsbach.2023b}. The reason is that it was extremely challenging, if not impossible, for research to systematically collect and holistically analyze content moderation decisions. In an attempt to foster transparency and accountability, the DSA has committed major social media platforms in the EU to make key information on their content moderation decisions accessible to the public. Here, we leverage the EU's Transparency Database -- a unique and previously unavailable data source -- to shed light on how major social media providers moderate user-generated content on their platforms. 

\textbf{Implications:} Our study implies differences and inconsistencies in the content moderation practices of major social media platforms. There are vast differences in the volume of content pieces that platforms moderate. For instance, TikTok performs more than 350 times more content moderation decisions per user than X/Twitter. It remains speculative whether platforms with higher content moderation volumes tend to encounter rule-breaking content more frequently or if they simply handle such content differently. However, our findings do indicate that different platforms tend to focus their content moderation efforts on different types of rule-breaking content. While all platforms consistently take action on violence and pornography/sexualized content, their approach to moderating other pertinent online harms, such as illegal/harmful speech and misinformation, varies considerably. For instance, on X/Twitter, illegal/harmful speech and misinformation are rarely moderated. Furthermore, the type of actions social media providers take against rule-breaking content varies across platforms. While most platforms frequently remove rule-breaking content, others are more inclined to reduce its visibility. Altogether, our study suggests that social media platforms interpret their obligations under the DSA differently -- resulting in a fragmented outcome that the DSA is meant to avoid. These findings have important implications for regulators, which may be inclined to clarify existing guidelines and/or lay out more specific rules that ensure common standards on how social media providers handle rule-breaking content on their platforms.

Furthermore, our findings contribute fresh insights to the ongoing debate (see \cite{Gillespie.2020}) regarding whether automation should have a role in social media content moderation or if this responsibility should predominantly rest with humans. Our results suggest that the majority of rule-breaking content is already identified and processed through automated methods rather than manual intervention. While many platforms (\eg, Facebook, Pinterest, and Instagram) employ a combination of automation and human moderation, others (\eg, TikTok) predominantly rely on fully automated content moderation. X/Twitter stands out as an exception, as it reports that it exclusively relies on non-automated means in its content moderation efforts. However, X/Twitter is also a platform that moderates a drastically lower volume of rule-breaking content compared to other platforms. Considering the immense scale of content moderation in the EU (more than 156 million instances within two months), this indicates there could be a necessity for some level of automation  to meet the regulatory obligations imposed on social media providers by the DSA.

\textbf{Limitations and future work:} Our work has several limitations, which provide promising opportunities for future research. First, our inferences are limited to the first {two} months after the introduction of the DSA Transparency Database in the EU. For this observation period, however, we analyze \emph{all} SoRs that were submitted by social media platforms. Second, content moderation efforts by social media platforms may evolve to a different steady-state due to growing experience, changes in functionality of the database, and clarification of rules by the EU. Future work may analyze how the patterns observed in this paper change over time. Third, more research is necessary to understand how content removal or visibility reduction affect on-platform user behavior. Fourth, it would be interesting to additionally analyze the source content that has been moderated (\eg, removed social media posts). However, this data is not publicly available from the EU. Notwithstanding these limitations, we believe that observing and understanding how social media providers moderate content on their platforms is the first step towards improving future policies, guidelines, and regulations. We hope that our early work inspires more research into improving transparency on content moderation on social media. 

\section{Conclusion}

The Digital Services Act (DSA) represents a major legislative framework in the EU that obligates large social media providers to publish clear and specific information whenever they remove or restrict access to certain content on their platforms. Due to this new level of transparency, we are, for the first time, able to empirically analyze real-world content moderation decisions of major social media platforms in the EU. Our empirical analysis based on more than 156 million SoRs suggests that there are  inconsistencies in how content moderation is carried out and how large social media platforms implement their obligations under the DSA. These findings hold important implications for regulators, who might be motivated to elucidate current guidelines or establish more specific regulations to ensure consistent standards for how social media platforms handle rule-breaking content on their platforms.


\bibliographystyle{ACM-Reference-Format-no-doi-abbrv}
\balance
\bibliography{literature}

\end{document}